\documentclass[RNAAS]{aastex62}

\begin{document}

\title{The Visualization and Measurements of Mass Functions with LEGO \footnote{Published in 2020, Research Notes of the AAS, 4, 134. 
Please see Kautsch et al. 2021, European Journal of Physics, 42, 035605, for more details and a discussion of the origin of mass functions.}}

\correspondingauthor{Stefan Kautsch}
\email{skautsch@nova.edu}

\author{Kyle K. Hansotia}
\altaffiliation{}
\affiliation{Nova Southeastern University\\
3301 College Ave.\\
Fort Lauderdale, FL 33314}

\author{Stefan J. Kautsch}
\altaffiliation{}
\affiliation{Nova Southeastern University\\
3301 College Ave.\\
Fort Lauderdale, FL 33314}

\begin{abstract}
We intend to promote the concept of mass functions for interdisciplinary science education and communication. 
A mass function characterizes the
frequency distribution of the masses of objects in the Universe. 
We present an experiment to demonstrate this concept to a diverse audience of science students, using popular LEGO toys.
We discovered that the LEGO mass function is surprisingly similar compared to mass functions of astronomical objects such as stars and galaxies.
\end{abstract} 

\keywords{Astronomy data visualization (1968), Stellar mass function (1612), Initial mass function (796), Interdisciplinary astronomy (804), Pareto distribution (1897), Mass ratio (1012)}

\section{Introduction} \label{intro}
We present an activity to visualize the concept of mass functions for the astronomy, physics, and general science curriculum in higher education, and public outreach. A mass function describes the frequency distribution of the masses of many objects. Mass functions can be frequently observed at ensembles of cosmic objects such as planets, stars, galaxies, dark halos, etc. Therefore, being aware of the concept of mass functions is fundamental to understand mass distribution and structure formation in the Universe. However, this concept is barely discussed in textbooks in astronomy and astrophysics for introductory courses, as well as in physics. It is even lesser known in other sciences. Only a few advanced, general astronomy textbooks (e.g., \citet{kut03}, \citet{co17}) cover the mass function of stars.

A simple mass function is the mathematical model $f(m)$ that fits the number of objects ($n$) in different ranges of mass ($m, m+dm$):
\begin{equation}
\label{mf}
\frac{d n}{d m}=f(m)=k m^{\alpha}
\end{equation}
$k$ is a coefficient, aka the constant. $\alpha$ is the power index, i.e., the slope of the power function. This type of function is also known as power law or scaling law. It shows that the number frequency of massive objects is much lower than the number of objects with smaller masses. Moreover, the proportion of high-to-low mass objects is constant.

In astronomy, mass functions are most commonly applied to stars since \citet{sal55}. The aim of these studies is to find the initial mass function (IMF, see the reviews in \citet{cor05}, and by \citet{kro13}, \citet{kru14}, \citet{hop18}, and references therein) and the present-day mass function (e.g., \citet{sca86}, \citet{cha03}, \citet{bov17}, \citet{sol19}), i.e., the proportion of stars of various masses and within a unit of volume at the moment of their birth or at present, respectively. Moreover, mass functions can be also found for other objects like galaxies, e.g., \citet{mof16}, and galaxy halos, e.g., \citet{pre74}, \citet{pen19}.

\citet{bin07} attempted to create a unified mass function based on an idea by \citet{zwi42} to show how cosmic objects are universally linked. They studied nearly all mass hierarchies in space, which include 36 orders of magnitude in mass, i.e., asteroids, planets, stars and their remnants, open and globular star clusters, molecular gas and dust clouds, galaxies, galaxy groups and clusters, and even simulated cold dark matter halos. They found that the mass distribution of these objects roughly follows a universal mass function of form $f(m) \propto m^{-2}$. This means that the proportion of high-to-low mass objects is always the same, no matter what kind of objects in the cosmos are considered. E.g., for stellar objects, it means that for each solar mass star, four stars with 1/2 the mass of the Sun exist, and 16 stars with 1/4 of the mass of the Sun exist.

Our intention of this work is to promote the concept of mass functions to be an integrated component of modern science education. We focus on the visualization of this concept using popular LEGO bricks. The choice of this toy enables easy visual understanding and application for college courses (life and online), laboratory courses, and independent studies.

\section{Methods} \label{methods}

Many LEGO sets contain a large amount of low-mass pieces and only a few massive bricks. This makes this toy ideal to show mass frequency distributions. We decided to use the set \href{https://www.lego.com/en-us/product/garmadon-garmadon-garmadon-70656}{\emph{LEGO 70656 NINJAGO garmadon, Garmadon, GARMADON!}}, because it contains a shark model like our university's mascot.

\begin{figure}[h!]
\begin{center}
\includegraphics[scale=0.57,angle=0]{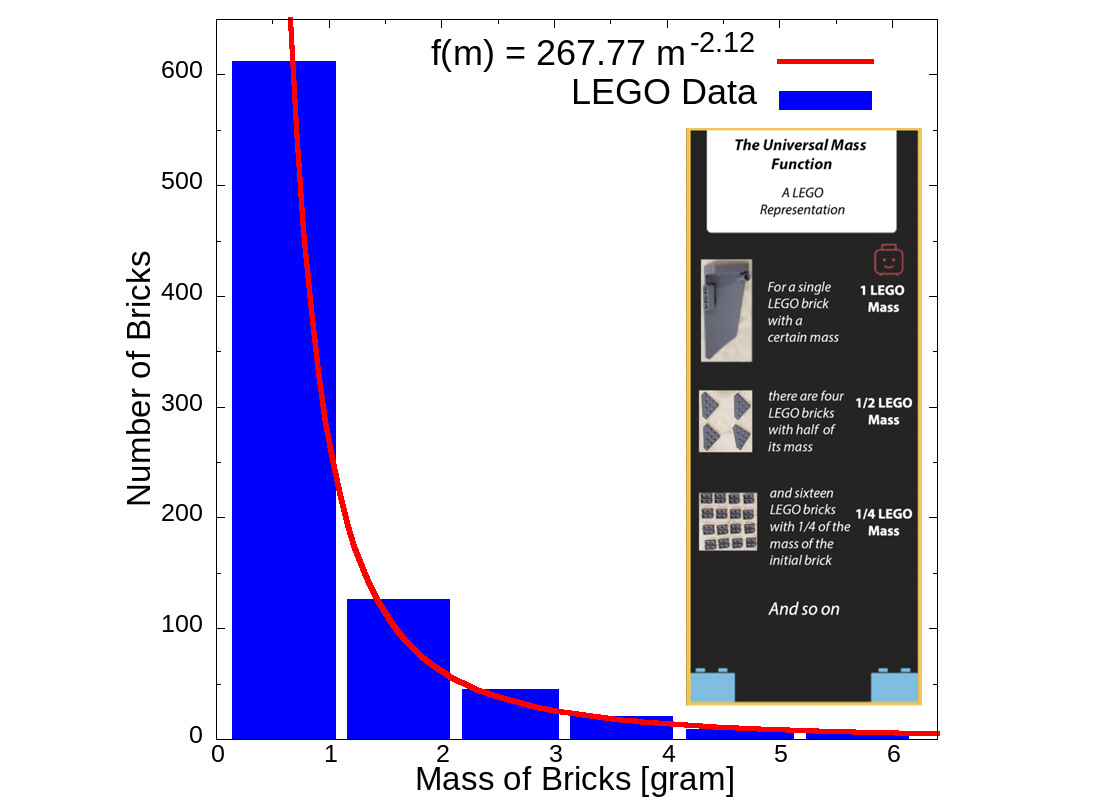}
\caption{The histogram shows the result of our measurements and fitting procedure. The LEGO bricks are sorted into six bins of increasing mass (blue), and fitted with the mass function power law (red line). The fitting results (slope and constant) are indicated in the legend. The graphic inlay shows the proportionality of the mass distribution of the LEGO bricks from our set. \label{fig1}}
\end{center}
\end{figure}

All pieces of this LEGO set were individually massed. Those bricks were then distributed into six equal-sized mass bins and counted. The number of bin intervals was naturally limited due to low numbers of massive bricks. These data were fit with the power law function of Eq. \ref{mf}. A nonlinear least-squares Marquardt-Levenberg fitting algorithm in gnuplot \citep{will2} was used to find the free parameters, i.e., the slope and the constant. Fig. \ref{fig1} shows the mass function as a line in a histogram with the mass of the bricks in grams in the bins on the x-axis and the numbers of bricks on the y-axis. The best-fit result for the slope is $\alpha = - 2.12 \pm 0.15$, and for the coefficient, it is $k = 267.77 \pm 1.20$.

\section{Results and Discussion} \label{results}

This simple experiment enables scientists and educators to inspire the audience in understanding the concept of mass functions. The quantitatively crucial part in understanding the mass function is the slope. This slope $\alpha$ describes how fast the numbers of objects increase with decreasing mass, thus, $\alpha$ determines the shape of the curve. Typical values of the slopes of stellar IMFs converge at values around $-2$ (e.g., \citet{sal55}, \citet{kro01}, \citet{cha03}, or \citet{wei13}). Also, \citet{bin07} found a slope of $\alpha = -2$ for their combined, universal mass function. It is very surprising that the slope of our LEGO mass function has a very similar value (i.e., $\alpha = -2.12$). This might be a fortunate coincidence of this particular set. However, it shows that LEGO is ideal to visualize this concept and should spark a classroom discussion where else this mass distribution behavior can be found in nature and how to explain it.

The physical origin of individual mass functions is highly debated (for the IMF see, e.g., \citet {cha03}, \citet {off16}). It is even lesser known why a universal mass function exists, which ranges from planetary bodies to galaxy clusters. \citet {bin07} point out that the formation scenarios of each object ensemble (planets vs. stars vs. galaxies, etc.) are totally different (bottom-up versus top-down). \citet{asc16}, \citet{asc18} and references therein explain in a comprehensive review that power law-like distributions can be observed in much more than the distribution of mass, e.g., structure distribution of stellar flares, black hole objects, planetary surface geometry, galactic structures, Pareto distributions in social science, just to name a few. They conclude that these distribution functions follow the concept of self-organizing critical processes. Applied to the mass function it could mean the ability of complex systems to self organize on many different mass scales in the Universe.

Therefore, we conclude that the concept of mass functions should be an integral part of science education. It will inspire students to learn more about the mysteries in our Universe and how the cosmos works. The experiential learning and teaching of mass functions using LEGO bricks can be applied in many creative ways because it is an excellent educational tool to visualize complex concepts.

\section{Acknowledgments} \label{ack}

This project was funded by Nova Southeastern University's President's Faculty Research and Development Grant 335510. We would like to thank Prof. Dr. B. Binggeli (University of Basel, Switzerland), Prof. Dr. D. Castano (Nova Southeastern University, U.S.A.), and Prof. Dr. D. Veras (University of Warwick, U.K.) for assisting us in the work process through intellectual conversations. The graph was made with \href{http://www.gnuplot.info}{gnuplot}. \href{https://www.adobe.com/products/photoshop.html}{Adobe Photoshop} was used to combine the \href{http://www.gnuplot.info}{gnuplot} histogram with the LEGO infographic. LEGO is a trademark of \href{https://www.lego.com/en-us/aboutus/lego-group/}{The LEGO Group}.

\end{document}